\documentclass[twocolumn, showpacs, showkeys, amsmath, amssymb,notitlepage, superscriptaddress,reprint,floatfix]{revtex4-1}

\usepackage[normalem]{ulem} 	
\usepackage[usenames,dvipsnames]{color}
\usepackage{upgreek}
\usepackage{siunitx}
\usepackage{graphicx}

\begin{document}

\title{Transmission-line resonators for the study of individual two-level tunneling systems}

\author{Jan David Brehm}
	\affiliation{Physikalisches Institut, Karlsruhe Institute of Technology, 76131 Karlsruhe, Germany}
\author{Alexander Bilmes}
	\affiliation{Physikalisches Institut, Karlsruhe Institute of Technology, 76131 Karlsruhe, Germany}
\author{Georg Weiss}
	\affiliation{Physikalisches Institut, Karlsruhe Institute of Technology, 76131 Karlsruhe, Germany}
\author{Alexey V. Ustinov}
	\affiliation{Physikalisches Institut, Karlsruhe Institute of Technology, 76131 Karlsruhe, Germany}
	\affiliation{Russian Quantum Center, National University of Science and Technology MISIS, Moscow 119049, Russia}
\author{J\"urgen Lisenfeld}
	\affiliation{Physikalisches Institut, Karlsruhe Institute of Technology, 76131 Karlsruhe, Germany}

\date{\today}

\begin{abstract}
Parasitic two-level tunneling systems (TLS) emerge in amorphous dielectrics and constitute a serious nuisance for various microfabricated devices, where they act as a source of noise and decoherence. Here, we demonstrate a new test bed for the study of TLS in various materials which provides access to properties of individual TLS as well as their ensemble response. We terminate a superconducting transmission-line resonator with a capacitor that hosts TLS in its dielectric. By tuning TLS via applied mechanical strain, we observe the signatures of individual TLS strongly coupled to the resonator in its transmission characteristics and extract the coupling components of their dipole moments and energy relaxation rates. The strong and well-defined coupling to the TLS bath results in pronounced resonator frequency fluctuations and excess phase noise, through which we can study TLS ensemble effects such as spectral diffusion, and probe theoretical models of TLS interaction.

\end{abstract}

\maketitle 

\setlength{\parskip}{-0.25cm}


\section{Introduction}
With the advent of experimental techniques to observe and manipulate the quantum states of superconducting circuits such as quantum bits and microwave resonators, two-level tunnelling systems (TLS) have been rediscovered as a major source of noise and decoherence. TLS are believed to originate in the tunnelling of an atomic entity between two nearby positions in a disordered material, giving rise to a microscopic elastic and, if charged, electric dipole which couple to electric fields and mechanical strain \cite{Phillips1972, Anderson1972}. In recent years, TLS which reside in amorphous dielectrics such as surface oxides or insulating coatings were increasingly recognized as important performance-limiting culprits in various systems, ranging from micromechanical resonators (MEMS) through kinetic inductance photon detectors (MKIDs) to interferometer mirrors used in gravitational wave detection~\cite{mems, mkids,Fejer16}.
After it was recognized that TLS within the tunnel barrier of a Josephson junction were the origin of avoided level crossings observed in spectroscopy of a superconducting phase qubit~\cite{Martinis:PRL:2005}, this type of device was utilized to achieve quantum state control and readout of individual TLS~\cite{TLSOscis,Lisenfeld:PRL:2010}. In such experiments, the ability to tune TLS via an applied mechanical strain~\cite{Grabovskij12} has proven to be particularly useful for studying TLS interactions~\cite{Lisenfeld2015} and decoherence~\cite{LisenfeldSciRep,Bilmes2016}.\\

\begin{figure}[b!]
	\includegraphics[width=8.7cm]{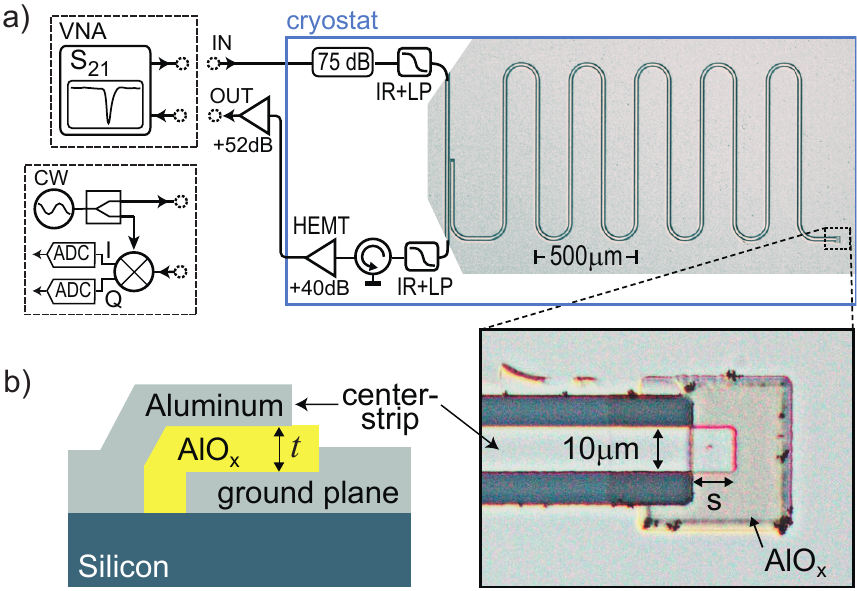}
	\caption{a) Schematic of the experimental setup and photograph of a sample, which is a $\frac{\lambda}{2}$-resonator terminated by an Al/AlO$_x$/Al capacitor as shown in the inset magnification. The dielectric volume of the capacitor is \SI{10}{\micro\metre}$\cdot s \cdot t$, with $t=50$ nm the AlO$_x$-thickness and lengths $s=\{5, 10, 15\}$\SI{}{\micro\metre}. The other end is capacitively coupled to a transmission line equipped with infrared (IR) and low-pass (LP) filters, isolators, a HEMT-amplifier at 4.2 K and additional room-temperature amplifiers. Measurements are done using either a network analyzer (VNA) or a continuous-wave (CW) microwave source and an IQ-mixer, whose I and Q outputs are sampled using fast analog to digital converters (ADC). b) Sketch of a capacitor cross-section.}\label{fig1}
\end{figure}

\begin{figure*}[t!]
	\includegraphics[width=\textwidth]{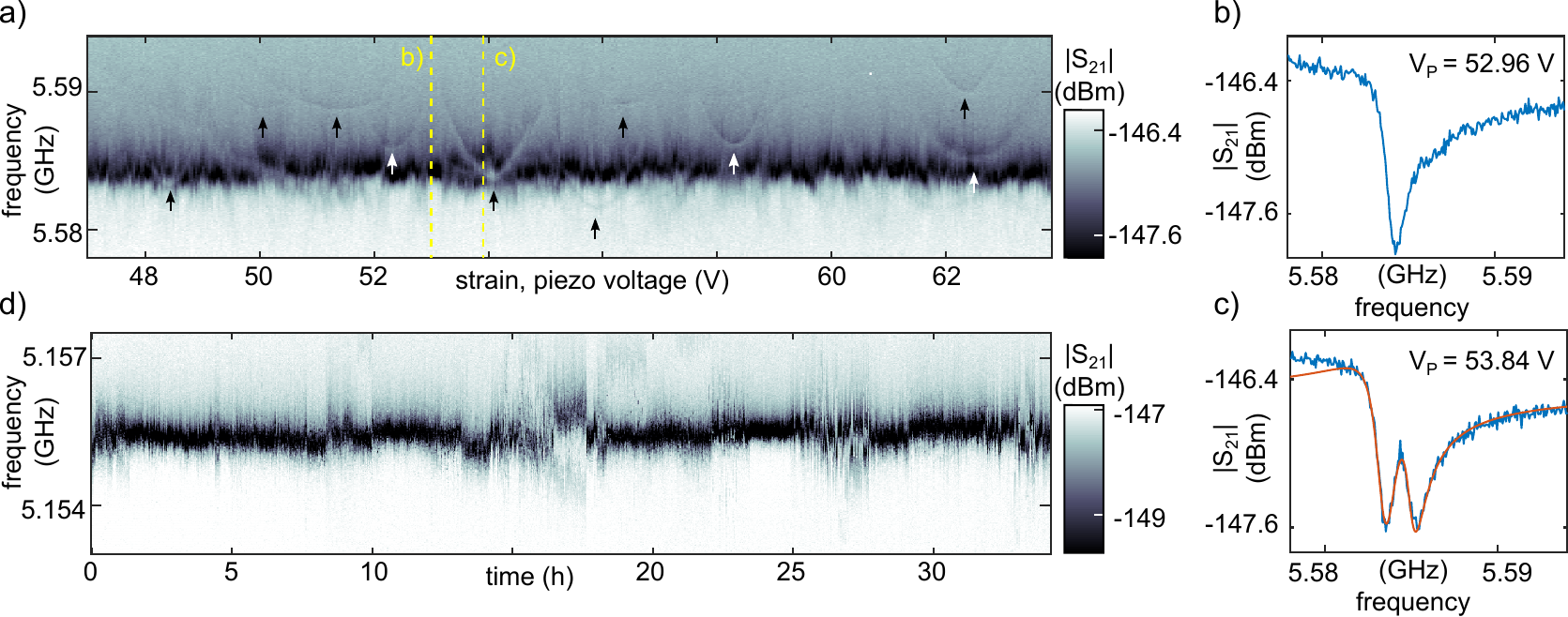}
	\caption{(a) Transmission $|S_{21}|$ (color-coded) of resonator 3 vs. frequency (vertical axis) and mechanical strain (in units of the applied piezo voltage, horizontal axis). Interaction with strongly coupled TLS is observed by hyperbolic traces when their resonance frequencies are strain-tuned through their symmetry points (indicated by arrows). Vertical dashed lines correspond to the cross-sections shown in b) and c). b) Typical resonator transmission vs. frequency. c) The transmission amplitude shows a double-dip feature due to a strongly coupled TLS near resonance with the resonator. The solid red line is a fit to theory, resulting in the TLS' decoherence rate, coupling strength and gap energy. d) Transmission of resonator 1 as a function of time at constant mechanical strain. Spectral TLS diffusion is observed by pronounced resonator frequency fluctuations, displaying steps and periods of strong telegraphic switching as well as slow drifts.}
	\label{fig2}
\end{figure*}
Employing qubits to study TLS is so far limited to addressing TLS in tunnel junction barriers, which for technical reasons are almost exclusively fabricated from thin ($\approx$ 2\,nm) layers of amorphous aluminum oxide, precluding studies on TLS in different materials. Only recently, strong coupling to individual TLS was observed using a superconducting lumped element LC-resonator which featured electric field tuning of TLS~\cite{OsbornTLSSpec}. Here, we present an alternative approach of addressing single TLS: We terminate a $\lambda$/2 transmission line resonator with an overlap capacitor that hosts TLS in its dielectric, and we tune TLS by mechanical strain rather than by an electric field. 
This setup constitutes a circuit-QED system with the dynamics described by the Jaynes-Cummings model~\cite{Schoelkopf04}, in which the coupling strength between a TLS and the resonator is given by $g=(\Delta/E) p_\parallel\, |E_\mathrm{RMS}|$, where $p_\parallel$ is the component of the TLS' dipole moment that is parallel to the oscillating electric field of amplitude $|E_\mathrm{RMS}|$ in the capacitor, $E$ is the TLS' excitation energy, and $\Delta$ is the tunneling energy between states. The coupling gives rise to a hybridization of TLS with cavity states and lifts the degeneracy of the corresponding excitation manifold. Depending on the TLS-resonator detuning $\delta$, the system shows either a resonant behavior or state-dependent dispersive resonance shifts by $2 g^2 / \delta$ for large detuning ($\delta > g$)~\cite{Schoelkopf04}.

\section{Experiment}
A sketch of our experimental setup and a sample photograph is shown in Fig.~\ref{fig1}\,a). The resonators are optically patterned from an Al film to form coplanar transmission lines of length $\lambda/2$ at a desired resonance frequency range from $5$ to $6\,$GHz. One end of the resonator's center strip is designed to overlap with a segment of the Al ground plane onto which AlO$_\text{x}$ has been deposited, thus forming an Al/AlO$_\text{x}$/Al plate capacitor as shown in the magnified inset to Fig.~\ref{fig1}\,a) and sketched in Fig.~\ref{fig1}\,b). The amorphous AlO$_\text{x}$ is fabricated using anodic oxidation of the ground plane to a desired thickness of $t=50$\,nm. To vary the dielectric volume $V_\mathrm{diel}=10$\,\SI{}{\micro\metre}$\cdot\,t\,\cdot s$ in the capacitor, we modify the overlap length $s$, resulting in the parameters summarized in Table~\ref{table1} for the three tested resonators.\\

For readout, the other end of the resonator is capacitively coupled to a transmission line whose input part is heavily attenuated and filtered at low temperatures. The output signal passes through two isolators and a large-gain, low-noise HEMT amplifier at 4.2 K and two additional amplifiers at room temperature. The transmission $S_{21}$ is recorded as a function of frequency using a network analyzer, see Fig.~\ref{fig2}\,b) for a typical result. The mechanical strain is controlled via the voltage $V_\text{p}$ applied to a piezo actuator which slightly bends the sample chip as described in Ref.~\cite{Grabovskij12}, resulting in a strain field $\epsilon \approx (8\cdot 10^{-7}/\text{V})\cdot V_\text{p}$. All data in this work were acquired at a sample temperature of $35\,$mK.

\section{Results}

{\bf TLS spectroscopy} is performed by recording the resonator transmission $S_{21}$ as a function of the applied mechanical strain. The excitation energy of a TLS is given by $E=\sqrt{\text{{$\it{\Delta}$}} ^2+\varepsilon ^2}$, where $\it{\Delta}$ is the (constant) tunnelling energy between the states and $\varepsilon=2\gamma (\epsilon-\epsilon_0)$ is the asymmetry energy which scales with the TLS' deformation potential $\gamma$ and the effective strain field $\epsilon$~\cite{Phillips87}. This hyperbolic dependence of TLS resonance frequencies can be directly observed by changes in the resonator transmission if the system is driven at powers below the one-photon-regime. In this power range TLS saturation effects can be neglected \cite{Phillips87}. We note that  all here observed TLS signatures are in direct vicinity of the TLS' symmetry points ($\varepsilon<10\,$MHz) such that $\Delta/E\approx1$ is valid in good approximation. \\ 

Figure~\ref{fig2}\,a) gives an example of such a measurement, plotting $S_{21}$ as a function of the applied piezo voltage. In the shown strain range, we observed about ten hyperbolic TLS traces, whose origins are marked by arrows. Fits to these traces directly result in the static TLS parameters $\it{\Delta}$ and $\gamma$. For the deformation potentials $\gamma$, we find values ranging between 0.1 and 1 eV (see also supplementary material), which are consistent with TLS ensemble measurements obtained on bulk AlO$_\text{x}$~\cite{LisenfeldSciRep}, and with experiments on individual TLS in tunnel junction barriers~\cite{Grabovskij12}. From extended measurements over a wider strain range, one could hypothetically reconstruct the distribution of tunnelling energies  $\it{\Delta}$ in order to verify a central assumption of the standard TLS model~\cite{Anderson1972,Phillips87}. By counting the number of visible hyperbolas, we extract densities of strongly coupled TLS of $5 -7 \,$(\SI{}{\micro\metre}$^3\,$GHz)$^{-1}$. This is consistent with measurements of TLS densities in AlO$_\text{x}$ tunnel junction barriers~\cite{Martinis:PRL:2005} when taking into account that the electric field strength in typically $\approx$ 2 nm thin barriers is about a factor of 25 larger as compared to the field strength in the capacitors employed here.
\\

Figure~\ref{fig2}\,c) shows the absolute value of the microwave transmission $|S_{21}|$ when a coherent TLS is tuned in resonance with the resonator. The signal displays two minima which frequencies are separated by an amount $\propto \sqrt{g^2+(\delta/2)^2}$. 
By fitting this data to a model obtained with input-output theory (see supplementary material and reference \cite{inputoutputtheory5}), we extract coupling strengths $g/2\pi$ between $0.5$ and $1.0\,$MHz, and TLS coherence times ranging from $102$ to \SI{363}{\nano\second}.
To obtain the TLS' electric dipole component $p_\parallel$ from the measured splitting size $g$, we estimate the electric field strength in the capacitors using SPICE simulations and find values ranging between 2.3 and $7.4\,$Debye (see Supplementary Material for individual TLS results and details on data analysis).\\

\begin{table} [tb!]
	\footnotesize 
	\begin{tabular}{l c c c c c c c c}  
		\#   & s         & $V_\mathrm{diel}~$&$C$ & $f_\mathrm{res}$  & $Q_\mathrm{c}$ & $F \tan(\delta_0)$ & $P_\text{c}$ &$\beta$ \\
		     &(\SI{}{\micro\metre})  &(\SI{}{\micro\metre^3}) &(fF)		& (GHz)				  & $ (10^{4})$ & $ (10^{-4})$ & (dBm) &\\ 
		 \hline 
		1       & 5   & 	2.5 & 89 &5.15775	& 29.3	& 0.4 $\pm$ 0.1 &-145 &0.97 $\pm$ 0.12 \\
		2      & 10  & 	5.0	 &178 & 5.46320 & 28.8	& 2.2 $\pm$ 0.2 &-145.6& 1.19 $\pm$ 0.03\\
		3       & 20  & 	10	& 356& 5.58385 & 22.7	& 1.9 $\pm$ 0.3 & -142.7&1.23 $\pm$ 0.06\\
		\hline \end{tabular} \normalsize 
\caption{Parameters of three tested resonators. $s$ is the designed lateral capacitor size (see Fig.~\ref{fig1}), $C$ its capacitance and $V_\mathrm{diel}$ the corresponding volume of its AlO$_x$ dielectric. $f_\mathrm{res}$ is the measured resonance frequency. The values of the coupling quality factor $Q_\text{c}$, the unsaturated loss tangent $F \tan(\delta_0)$, the critical power $P_\text{c}$ and the exponent $\beta$ are extracted from fits to the power-dependent resonator loss rate Eq.~(\ref{tlsloss}). The errors on $P_\text{c}$ are smaller than $10^{-3}\,$ dBm and therefore not shown.}
\label{table1}

\end{table}
{\bf Spectral diffusion} denotes fluctuations of the TLS' excitation energy $E$ due to their non-resonant interaction with neighboring thermal TLS~\cite{Phillips87}. The transition energies of such thermal TLS are below $k_\text{B}T$ such that they experience random state switching under thermal fluctuations and therefore change the local strain and electric field. As a consequence, the detuning $\delta$ between coherent TLS and the resonator depends on time which gives rise to resonator noise induced by the fluctuating dispersive shift ~\cite{Gao07,FaoroResonators}. In our system, we observe pronounced frequency and phase fluctuations in the resonator due to its strong coupling to TLS in the capacitor. As an example, in Fig.~\ref{fig2}\,d) we present a series of transmission curves measured at a constant mechanical strain over a course of 35 hours. One can distinguish rather quiet periods characterized by small losses and fluctuations, times with slow drifts (due to a coupling to an ensemble of thermal TLS) and also sudden steps in frequency, as well as periods with pronounced telegraphic switching between two resonance frequencies (due to a coupling to single thermal TLS).\\

{\bf Power dependence.} 
It is well established that TLS residing on surface oxides of coplanar resonators give rise to a power dependent dielectric loss rate  $\tan(\delta)$ (or the inverse internal quality factor $Q_\text{i}$)~\cite{TLSloss,TLSloss2,TLSloss3}. Once the internal resonator power $P_\text{i}$ exceeds a critical value $P_\text{c}$, the loss decreases because resonant TLS start to become saturated and act no longer as an effective photon sink~\cite{Schickfus77}. The loss rate is typically modelled as
\begin{equation}
\tan(\delta)=Q_\mathrm{i}^{-1}=\frac{F\tan(\delta_0)\tanh(\frac{\hbar \omega}{2k_\mathrm{B}T})}{\sqrt{1+\left(P_\mathrm{i}/P_\mathrm{c}\right)^\beta}}+\tan(\delta_r),
\label{tlsloss}
\end{equation}
where $\tan(\delta_r)$ is a residual loss rate due to eg. radiative, vortex or quasiparticle losses. $F \tan(\delta_0)$ is the dielectric loss rate of aluminum oxide dressed by a filling factor $F$ which takes into account the participation ratio of the lossy material. Furthermore the exponent $\beta$ is introduced to account for inhomogeneous field distributions depending on the specific geometry of the device~\cite{TLSloss3,TLSloss}.
\\ 
In our system, we distinguish two baths of TLS: those in the surface oxide of the resonator, and those in the dielectric of the overlap capacitor. We expect the latter to dominate the low-power resonator losses, because their coupling strength is orders of magnitude larger. Comparing the electric field energies in the two TLS-hosting volumes, we find participation ratios of surface and capacitor TLS $p_\text{surf}/p_\text{cap}\approx 2-9\cdot10^{-4}$, depending on the capacitor sizes.\\ 

\begin{figure}[tb]
\includegraphics[width=8.7cm]{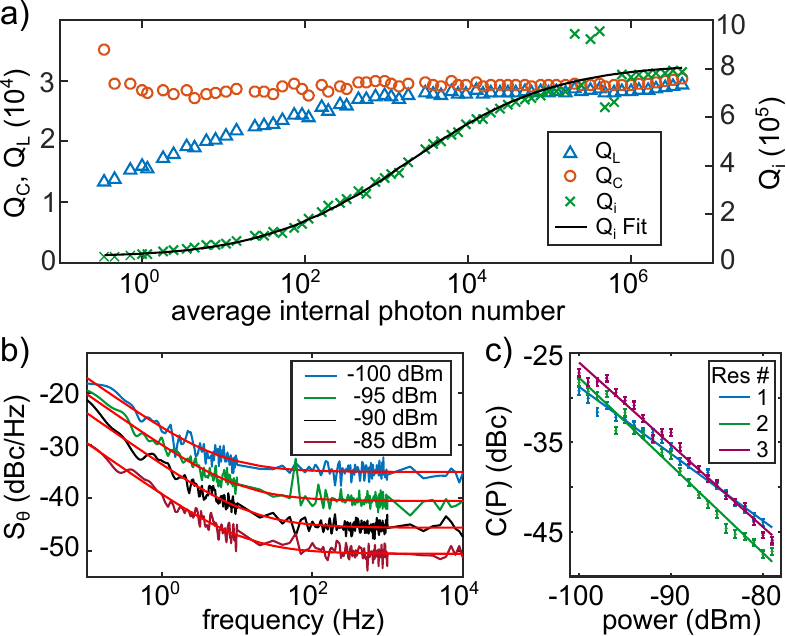}
\caption{a) Power dependence of quality factors extracted from resonator 1 at $35\,$mK. For powers higher than the single-photon regime $(>-142\,\text{dBm})$, the internal Q-factor $Q_\text{i}$ increases significantly from a minimal value of $2.1\cdot10^4$ due to TLS saturation as expected from the standard tunnelling model. A fit to Eq.~(\ref{tlsloss}) (solid line) results in the parameters $F \tan(\delta_0)$, $P_\text{c}$ and $\beta$ given in Table~\ref{table1}. 
b) Power dependence of the phase noise spectral density $S_\theta$ and c) its low frequency 1/f-noise amplitude $C(P)$. 
}
\label{fig3}
\end{figure}
The extracted dependence of the internal, coupling and loaded Q-factors ($Q_\text{i}$, $Q_\text{C}$, $Q_\text{L}$) on average photon number is shown for resonator 1 in Fig.~\ref{fig3}\,a). The saturation of TLS with increasing power can clearly be seen in the increase of the internal Q-factor. A fit to Eq.~(\ref{tlsloss}) provides the parameters listed in table \ref{table1}. The extracted critical powers $P_\text{C}$ are for all resonators in the one-photon regime, which indicates the presence of single strongly coupled TLS. We estimate the uncertainty in the internal photon number to a factor of $\approx 2 - 3$ since we had no possibility to calibrate the attenuation in the cryostat's transmission line in-situ. \\

{\bf Resonator noise.}
In previous work it has been shown that TLS produce low-frequency 1/f-phase or frequency noise in resonators \cite{Gao07}, and an anomalous scaling of the phase noise amplitude with temperature has been found which contradicts the predictions of the Standard Tunnelling Model \cite{burnett_evidence_2014, 1/fnoise}. Recently, two theories have incorporated TLS-TLS interactions in form of spectral diffusion to explain the anomalous scaling \cite{1/fnoise_burin,1/fnoise_faoro}.\\
In order to test these predictions for our system, we use a homodyne detection setup depicted in Fig.~\ref{fig1}a). The phase noise spectral density $S_\theta(f)$ is extracted as described in reference \cite{Gao07}. Figure \ref{fig3}b) shows $S_\theta(f)$ for different input powers. To split the low frequency 1/f-contribution and the constant noise floor, $S_\theta(f)$ is fitted to $C(P)/f+D$ , where $C(P)$ is the 1/f-noise amplitude. Its scaling with power is depicted in Fig. \ref{fig3}c) for the three resonators. For our resonators we extract the power-laws $C(P)\propto 1/P^{\beta_N}$ with $\beta_N=0.75,\,1.00,\,0.92$ which point towards a model of weakly interacting TLS \cite{1/fnoise_burin} and agree well with previously measured data \cite{1/fnoise}.\\

\section{Conclusions}
We have presented a new method to study the properties of both single TLS and TLS ensembles by coupling an overlap capacitor with a TLS-hosting dielectric to a superconducting transmission line resonator. By applying mechanical strain to tune strongly coupled TLS near resonance with the resonator, we perform TLS spectroscopy and extract their individual dipole moments, deformation potentials, tunnel energies and coherence times. Our system features strong coupling to a defined set of TLS and facilitates measurements of spectral diffusion and phase noise. This method can be applied to study TLS in various materials, while requiring only modest fabrication efforts and standard experimental techniques. It can become a useful tool to characterize materials employed in quantum circuits and to obtain a better understanding of the microscopic origin of TLS.\\

\noindent\textbf{Supplementary Material}
See supplementary material for additional information about individual TLS parameters and input-output theory.\\

\noindent\textbf{Acknowledgements.}
This work was funded by the Deutsche Forschungsgesellschaft (DFG), grant LI2446/1-1 and partially supported by the Ministry of Education and Science of Russian Federation in the framework of Increase Competitiveness Program of the NUST MISIS (contracts no. K2-2015-002 and K2-2016-063). Alexander Bilmes acknowledges support from the Helmholtz International Research School for Teratronics (HIRST) and the Landesgraduiertenf\"orderung-Karlsruhe (LGF).

\bibliography{TLS_CPW_bib}

\begin{thebibliography}{26}%
\makeatletter
\providecommand \@ifxundefined [1]{%
 \@ifx{#1\undefined}
}%
\providecommand \@ifnum [1]{%
 \ifnum #1\expandafter \@firstoftwo
 \else \expandafter \@secondoftwo
 \fi
}%
\providecommand \@ifx [1]{%
 \ifx #1\expandafter \@firstoftwo
 \else \expandafter \@secondoftwo
 \fi
}%
\providecommand \natexlab [1]{#1}%
\providecommand \enquote  [1]{``#1''}%
\providecommand \bibnamefont  [1]{#1}%
\providecommand \bibfnamefont [1]{#1}%
\providecommand \citenamefont [1]{#1}%
\providecommand \href@noop [0]{\@secondoftwo}%
\providecommand \href [0]{\begingroup \@sanitize@url \@href}%
\providecommand \@href[1]{\@@startlink{#1}\@@href}%
\providecommand \@@href[1]{\endgroup#1\@@endlink}%
\providecommand \@sanitize@url [0]{\catcode `\\12\catcode `\$12\catcode
  `\&12\catcode `\#12\catcode `\^12\catcode `\_12\catcode `\%12\relax}%
\providecommand \@@startlink[1]{}%
\providecommand \@@endlink[0]{}%
\providecommand \url  [0]{\begingroup\@sanitize@url \@url }%
\providecommand \@url [1]{\endgroup\@href {#1}{\urlprefix }}%
\providecommand \urlprefix  [0]{URL }%
\providecommand \Eprint [0]{\href }%
\providecommand \doibase [0]{http://dx.doi.org/}%
\providecommand \selectlanguage [0]{\@gobble}%
\providecommand \bibinfo  [0]{\@secondoftwo}%
\providecommand \bibfield  [0]{\@secondoftwo}%
\providecommand \translation [1]{[#1]}%
\providecommand \BibitemOpen [0]{}%
\providecommand \bibitemStop [0]{}%
\providecommand \bibitemNoStop [0]{.\EOS\space}%
\providecommand \EOS [0]{\spacefactor3000\relax}%
\providecommand \BibitemShut  [1]{\csname bibitem#1\endcsname}%
\let\auto@bib@innerbib\@empty
\bibitem [{\citenamefont {Phillips}(1972)}]{Phillips1972}%
  \BibitemOpen
  \bibfield  {author} {\bibinfo {author} {\bibfnamefont {W.~A.}\ \bibnamefont
  {Phillips}},\ }\href {\doibase 10.1007/BF00660072} {\bibfield  {journal}
  {\bibinfo  {journal} {Journal of Low Temperature Physics}\ }\textbf {\bibinfo
  {volume} {7}},\ \bibinfo {pages} {351} (\bibinfo {year} {1972})}\BibitemShut
  {NoStop}%
\bibitem [{\citenamefont {Anderson}\ \emph {et~al.}(1972)\citenamefont
  {Anderson}, \citenamefont {Halperin},\ and\ \citenamefont
  {Varma}}]{Anderson1972}%
  \BibitemOpen
  \bibfield  {author} {\bibinfo {author} {\bibfnamefont {P.~W.}\ \bibnamefont
  {Anderson}}, \bibinfo {author} {\bibfnamefont {B.~I.}\ \bibnamefont
  {Halperin}}, \ and\ \bibinfo {author} {\bibfnamefont {C.}~\bibnamefont
  {Varma}},\ }\href@noop {} {\bibfield  {journal} {\bibinfo  {journal} {Philos.
  Mag.}\ }\textbf {\bibinfo {volume} {25}},\ \bibinfo {pages} {1} (\bibinfo
  {year} {1972})}\BibitemShut {NoStop}%
\bibitem [{\citenamefont {Mohanty}\ \emph {et~al.}(2002)\citenamefont
  {Mohanty}, \citenamefont {Harrington}, \citenamefont {Ekinci}, \citenamefont
  {Yang}, \citenamefont {Murphy},\ and\ \citenamefont {Roukes}}]{mems}%
  \BibitemOpen
  \bibfield  {author} {\bibinfo {author} {\bibfnamefont {P.}~\bibnamefont
  {Mohanty}}, \bibinfo {author} {\bibfnamefont {D.~A.}\ \bibnamefont
  {Harrington}}, \bibinfo {author} {\bibfnamefont {K.~L.}\ \bibnamefont
  {Ekinci}}, \bibinfo {author} {\bibfnamefont {Y.~T.}\ \bibnamefont {Yang}},
  \bibinfo {author} {\bibfnamefont {M.~J.}\ \bibnamefont {Murphy}}, \ and\
  \bibinfo {author} {\bibfnamefont {M.~L.}\ \bibnamefont {Roukes}},\ }\href
  {\doibase 10.1103/PhysRevB.66.085416} {\bibfield  {journal} {\bibinfo
  {journal} {Phys. Rev. B}\ }\textbf {\bibinfo {volume} {66}},\ \bibinfo
  {pages} {085416} (\bibinfo {year} {2002})}\BibitemShut {NoStop}%
\bibitem [{\citenamefont {Janssen}\ \emph {et~al.}(2013)\citenamefont
  {Janssen}, \citenamefont {Baselmans}, \citenamefont {Endo}, \citenamefont
  {Ferrari}, \citenamefont {Yates}, \citenamefont {Baryshev},\ and\
  \citenamefont {Klapwijk}}]{mkids}%
  \BibitemOpen
  \bibfield  {author} {\bibinfo {author} {\bibfnamefont {R.~M.~J.}\
  \bibnamefont {Janssen}}, \bibinfo {author} {\bibfnamefont {J.~J.~A.}\
  \bibnamefont {Baselmans}}, \bibinfo {author} {\bibfnamefont {A.}~\bibnamefont
  {Endo}}, \bibinfo {author} {\bibfnamefont {L.}~\bibnamefont {Ferrari}},
  \bibinfo {author} {\bibfnamefont {S.~J.~C.}\ \bibnamefont {Yates}}, \bibinfo
  {author} {\bibfnamefont {A.~M.}\ \bibnamefont {Baryshev}}, \ and\ \bibinfo
  {author} {\bibfnamefont {T.~M.}\ \bibnamefont {Klapwijk}},\ }\href {\doibase
  10.1063/1.4829657} {\bibfield  {journal} {\bibinfo  {journal} {Appl. Phys.
  Lett.}\ }\textbf {\bibinfo {volume} {103}},\ \bibinfo {pages} {203503}
  (\bibinfo {year} {2013})}\BibitemShut {NoStop}%
\bibitem [{\citenamefont {Fejer}(2016)}]{Fejer16}%
  \BibitemOpen
  \bibfield  {author} {\bibinfo {author} {\bibfnamefont {M.~M.}\ \bibnamefont
  {Fejer}},\ }in\ \href {\doibase 10.1364/OIC.2016.MB.1} {\emph {\bibinfo
  {booktitle} {Optical Interference Coatings 2016}}}\ (\bibinfo  {publisher}
  {Optical Society of America},\ \bibinfo {year} {2016})\ p.\ \bibinfo {pages}
  {MB.1}\BibitemShut {NoStop}%
\bibitem [{\citenamefont {Martinis}\ \emph {et~al.}(2005)\citenamefont
  {Martinis}, \citenamefont {Cooper}, \citenamefont {McDermott}, \citenamefont
  {Steffen}, \citenamefont {Ansmann}, \citenamefont {Osborn}, \citenamefont
  {Cicak}, \citenamefont {Oh}, \citenamefont {Pappas}, \citenamefont
  {Simmonds},\ and\ \citenamefont {Yu}}]{Martinis:PRL:2005}%
  \BibitemOpen
  \bibfield  {author} {\bibinfo {author} {\bibfnamefont {J.~M.}\ \bibnamefont
  {Martinis}}, \bibinfo {author} {\bibfnamefont {K.~B.}\ \bibnamefont
  {Cooper}}, \bibinfo {author} {\bibfnamefont {R.}~\bibnamefont {McDermott}},
  \bibinfo {author} {\bibfnamefont {M.}~\bibnamefont {Steffen}}, \bibinfo
  {author} {\bibfnamefont {M.}~\bibnamefont {Ansmann}}, \bibinfo {author}
  {\bibfnamefont {K.~D.}\ \bibnamefont {Osborn}}, \bibinfo {author}
  {\bibfnamefont {K.}~\bibnamefont {Cicak}}, \bibinfo {author} {\bibfnamefont
  {S.}~\bibnamefont {Oh}}, \bibinfo {author} {\bibfnamefont {D.~P.}\
  \bibnamefont {Pappas}}, \bibinfo {author} {\bibfnamefont {R.~W.}\
  \bibnamefont {Simmonds}}, \ and\ \bibinfo {author} {\bibfnamefont {C.~C.}\
  \bibnamefont {Yu}},\ }\href@noop {} {\bibfield  {journal} {\bibinfo
  {journal} {Phys. Rev. Lett.}\ }\textbf {\bibinfo {volume} {95}},\ \bibinfo
  {pages} {210503} (\bibinfo {year} {2005})}\BibitemShut {NoStop}%
\bibitem [{\citenamefont {Cooper}\ \emph {et~al.}(2004)\citenamefont {Cooper},
  \citenamefont {Steffen}, \citenamefont {McDermott}, \citenamefont {Simmonds},
  \citenamefont {Seongshik}, \citenamefont {Hite}, \citenamefont {Pappas},\
  and\ \citenamefont {Martinis}}]{TLSOscis}%
  \BibitemOpen
  \bibfield  {author} {\bibinfo {author} {\bibfnamefont {K.~B.}\ \bibnamefont
  {Cooper}}, \bibinfo {author} {\bibfnamefont {M.}~\bibnamefont {Steffen}},
  \bibinfo {author} {\bibfnamefont {R.}~\bibnamefont {McDermott}}, \bibinfo
  {author} {\bibfnamefont {R.}~\bibnamefont {Simmonds}}, \bibinfo {author}
  {\bibfnamefont {O.}~\bibnamefont {Seongshik}}, \bibinfo {author}
  {\bibfnamefont {D.~A.}\ \bibnamefont {Hite}}, \bibinfo {author}
  {\bibfnamefont {D.~P.}\ \bibnamefont {Pappas}}, \ and\ \bibinfo {author}
  {\bibfnamefont {J.~M.}\ \bibnamefont {Martinis}},\ }\href@noop {} {\bibfield
  {journal} {\bibinfo  {journal} {Phys. Rev. Lett.}\ }\textbf {\bibinfo
  {volume} {93}},\ \bibinfo {pages} {180401} (\bibinfo {year}
  {2004})}\BibitemShut {NoStop}%
\bibitem [{\citenamefont {Lisenfeld}\ \emph {et~al.}(2010)\citenamefont
  {Lisenfeld}, \citenamefont {M{\"u}ller}, \citenamefont {Cole}, \citenamefont
  {Bushev}, \citenamefont {Lukashenko}, \citenamefont {Shnirman},\ and\
  \citenamefont {Ustinov}}]{Lisenfeld:PRL:2010}%
  \BibitemOpen
  \bibfield  {author} {\bibinfo {author} {\bibfnamefont {J.}~\bibnamefont
  {Lisenfeld}}, \bibinfo {author} {\bibfnamefont {C.}~\bibnamefont
  {M{\"u}ller}}, \bibinfo {author} {\bibfnamefont {J.~H.}\ \bibnamefont
  {Cole}}, \bibinfo {author} {\bibfnamefont {P.}~\bibnamefont {Bushev}},
  \bibinfo {author} {\bibfnamefont {A.}~\bibnamefont {Lukashenko}}, \bibinfo
  {author} {\bibfnamefont {A.}~\bibnamefont {Shnirman}}, \ and\ \bibinfo
  {author} {\bibfnamefont {A.~V.}\ \bibnamefont {Ustinov}},\ }\href@noop {}
  {\bibfield  {journal} {\bibinfo  {journal} {Phys. Rev. Lett.}\ }\textbf
  {\bibinfo {volume} {105}},\ \bibinfo {pages} {230504} (\bibinfo {year}
  {2010})}\BibitemShut {NoStop}%
\bibitem [{\citenamefont {Grabovskij}\ \emph {et~al.}(2012)\citenamefont
  {Grabovskij}, \citenamefont {Peichl}, \citenamefont {Lisenfeld},
  \citenamefont {Weiss},\ and\ \citenamefont {Ustinov}}]{Grabovskij12}%
  \BibitemOpen
  \bibfield  {author} {\bibinfo {author} {\bibfnamefont {G.~J.}\ \bibnamefont
  {Grabovskij}}, \bibinfo {author} {\bibfnamefont {T.}~\bibnamefont {Peichl}},
  \bibinfo {author} {\bibfnamefont {J.}~\bibnamefont {Lisenfeld}}, \bibinfo
  {author} {\bibfnamefont {G.}~\bibnamefont {Weiss}}, \ and\ \bibinfo {author}
  {\bibfnamefont {A.~V.}\ \bibnamefont {Ustinov}},\ }\href@noop {} {\bibfield
  {journal} {\bibinfo  {journal} {Science}\ }\textbf {\bibinfo {volume}
  {338}},\ \bibinfo {pages} {232} (\bibinfo {year} {2012})}\BibitemShut
  {NoStop}%
\bibitem [{\citenamefont {Lisenfeld}\ \emph {et~al.}(2015)\citenamefont
  {Lisenfeld}, \citenamefont {Grabovskij}, \citenamefont {M\"uller},
  \citenamefont {Cole}, \citenamefont {Weiss},\ and\ \citenamefont
  {Ustinov}}]{Lisenfeld2015}%
  \BibitemOpen
  \bibfield  {author} {\bibinfo {author} {\bibfnamefont {J.}~\bibnamefont
  {Lisenfeld}}, \bibinfo {author} {\bibfnamefont {G.}~\bibnamefont
  {Grabovskij}}, \bibinfo {author} {\bibfnamefont {C.}~\bibnamefont
  {M\"uller}}, \bibinfo {author} {\bibfnamefont {J.}~\bibnamefont {Cole}},
  \bibinfo {author} {\bibfnamefont {G.}~\bibnamefont {Weiss}}, \ and\ \bibinfo
  {author} {\bibfnamefont {A.}~\bibnamefont {Ustinov}},\ }\href {\doibase
  10.1038/ncomms7182} {\bibfield  {journal} {\bibinfo  {journal} {Nat. Comm.}\
  }\textbf {\bibinfo {volume} {6}},\ \bibinfo {pages} {6182} (\bibinfo {year}
  {2015})}\BibitemShut {NoStop}%
\bibitem [{\citenamefont {Lisenfeld}\ \emph {et~al.}(2016)\citenamefont
  {Lisenfeld}, \citenamefont {Bilmes}, \citenamefont {Matityahu}, \citenamefont
  {Zanker}, \citenamefont {Marthaler}, \citenamefont {Schechter}, \citenamefont
  {Sch\"on}, \citenamefont {Shnirman}, \citenamefont {Weiss},\ and\
  \citenamefont {Ustinov}}]{LisenfeldSciRep}%
  \BibitemOpen
  \bibfield  {author} {\bibinfo {author} {\bibfnamefont {J.}~\bibnamefont
  {Lisenfeld}}, \bibinfo {author} {\bibfnamefont {A.}~\bibnamefont {Bilmes}},
  \bibinfo {author} {\bibfnamefont {S.}~\bibnamefont {Matityahu}}, \bibinfo
  {author} {\bibfnamefont {S.}~\bibnamefont {Zanker}}, \bibinfo {author}
  {\bibfnamefont {M.}~\bibnamefont {Marthaler}}, \bibinfo {author}
  {\bibfnamefont {M.}~\bibnamefont {Schechter}}, \bibinfo {author}
  {\bibfnamefont {G.}~\bibnamefont {Sch\"on}}, \bibinfo {author} {\bibfnamefont
  {A.}~\bibnamefont {Shnirman}}, \bibinfo {author} {\bibfnamefont
  {G.}~\bibnamefont {Weiss}}, \ and\ \bibinfo {author} {\bibfnamefont {A.~V.}\
  \bibnamefont {Ustinov}},\ }\href {\doibase 10.1038/srep23786} {\bibfield
  {journal} {\bibinfo  {journal} {Sci. Rep.}\ }\textbf {\bibinfo {volume}
  {6}},\ \bibinfo {pages} {23786} (\bibinfo {year} {2016})}\BibitemShut
  {NoStop}%
\bibitem [{\citenamefont {Bilmes}\ \emph {et~al.}(2017)\citenamefont {Bilmes},
  \citenamefont {Zanker}, \citenamefont {Heimes}, \citenamefont {Marthaler},
  \citenamefont {Sch\"on}, \citenamefont {Weiss}, \citenamefont {Ustinov},\
  and\ \citenamefont {Lisenfeld}}]{Bilmes2016}%
  \BibitemOpen
  \bibfield  {author} {\bibinfo {author} {\bibfnamefont {A.}~\bibnamefont
  {Bilmes}}, \bibinfo {author} {\bibfnamefont {S.}~\bibnamefont {Zanker}},
  \bibinfo {author} {\bibfnamefont {A.}~\bibnamefont {Heimes}}, \bibinfo
  {author} {\bibfnamefont {M.}~\bibnamefont {Marthaler}}, \bibinfo {author}
  {\bibfnamefont {G.}~\bibnamefont {Sch\"on}}, \bibinfo {author} {\bibfnamefont
  {G.}~\bibnamefont {Weiss}}, \bibinfo {author} {\bibfnamefont {A.~V.}\
  \bibnamefont {Ustinov}}, \ and\ \bibinfo {author} {\bibfnamefont
  {J.}~\bibnamefont {Lisenfeld}},\ }\href {\doibase 10.1103/PhysRevB.96.064504}
  {\bibfield  {journal} {\bibinfo  {journal} {Phys. Rev. B}\ }\textbf {\bibinfo
  {volume} {96}},\ \bibinfo {pages} {064504} (\bibinfo {year}
  {2017})}\BibitemShut {NoStop}%
\bibitem [{\citenamefont {Sarabi}\ \emph {et~al.}(2016)\citenamefont {Sarabi},
  \citenamefont {Ramanayaka}, \citenamefont {Burin}, \citenamefont
  {Wellstood},\ and\ \citenamefont {Osborn}}]{OsbornTLSSpec}%
  \BibitemOpen
  \bibfield  {author} {\bibinfo {author} {\bibfnamefont {B.}~\bibnamefont
  {Sarabi}}, \bibinfo {author} {\bibfnamefont {A.~N.}\ \bibnamefont
  {Ramanayaka}}, \bibinfo {author} {\bibfnamefont {A.~L.}\ \bibnamefont
  {Burin}}, \bibinfo {author} {\bibfnamefont {F.~C.}\ \bibnamefont
  {Wellstood}}, \ and\ \bibinfo {author} {\bibfnamefont {K.~D.}\ \bibnamefont
  {Osborn}},\ }\href {\doibase 10.1103/PhysRevLett.116.167002} {\bibfield
  {journal} {\bibinfo  {journal} {Phys. Rev. Lett.}\ }\textbf {\bibinfo
  {volume} {116}},\ \bibinfo {pages} {167002} (\bibinfo {year}
  {2016})}\BibitemShut {NoStop}%
\bibitem [{\citenamefont {Blais}\ \emph {et~al.}(2004)\citenamefont {Blais},
  \citenamefont {Huang}, \citenamefont {Wallraff}, \citenamefont {Girvin},\
  and\ \citenamefont {Schoelkopf}}]{Schoelkopf04}%
  \BibitemOpen
  \bibfield  {author} {\bibinfo {author} {\bibfnamefont {A.}~\bibnamefont
  {Blais}}, \bibinfo {author} {\bibfnamefont {R.-S.}\ \bibnamefont {Huang}},
  \bibinfo {author} {\bibfnamefont {A.}~\bibnamefont {Wallraff}}, \bibinfo
  {author} {\bibfnamefont {S.~M.}\ \bibnamefont {Girvin}}, \ and\ \bibinfo
  {author} {\bibfnamefont {R.~J.}\ \bibnamefont {Schoelkopf}},\ }\href
  {\doibase 10.1103/PhysRevA.69.062320} {\bibfield  {journal} {\bibinfo
  {journal} {Phys. Rev. A}\ }\textbf {\bibinfo {volume} {69}},\ \bibinfo
  {pages} {062320} (\bibinfo {year} {2004})}\BibitemShut {NoStop}%
\bibitem [{\citenamefont {Phillips}(1987)}]{Phillips87}%
  \BibitemOpen
  \bibfield  {author} {\bibinfo {author} {\bibfnamefont {W.~A.}\ \bibnamefont
  {Phillips}},\ }\href@noop {} {\bibfield  {journal} {\bibinfo  {journal} {Rep.
  Prog. Phys.}\ }\textbf {\bibinfo {volume} {50}},\ \bibinfo {pages} {1657}
  (\bibinfo {year} {1987})}\BibitemShut {NoStop}%
\bibitem [{\citenamefont {Sarabi}\ \emph {et~al.}(2015)\citenamefont {Sarabi},
  \citenamefont {Ramanayaka}, \citenamefont {Burin}, \citenamefont
  {Wellstood},\ and\ \citenamefont {Osborn}}]{inputoutputtheory5}%
  \BibitemOpen
  \bibfield  {author} {\bibinfo {author} {\bibfnamefont {B.}~\bibnamefont
  {Sarabi}}, \bibinfo {author} {\bibfnamefont {A.~N.}\ \bibnamefont
  {Ramanayaka}}, \bibinfo {author} {\bibfnamefont {A.~L.}\ \bibnamefont
  {Burin}}, \bibinfo {author} {\bibfnamefont {F.~C.}\ \bibnamefont
  {Wellstood}}, \ and\ \bibinfo {author} {\bibfnamefont {K.~D.}\ \bibnamefont
  {Osborn}},\ }\href {\doibase 10.1063/1.4918775} {\bibfield  {journal}
  {\bibinfo  {journal} {Appl. Phys. Lett.}\ }\textbf {\bibinfo {volume}
  {106}},\ \bibinfo {pages} {172601} (\bibinfo {year} {2015})}\BibitemShut
  {NoStop}%
\bibitem [{\citenamefont {Gao}\ \emph {et~al.}(2007)\citenamefont {Gao},
  \citenamefont {Zmuidzinas}, \citenamefont {Mazin}, \citenamefont {LeDuc},\
  and\ \citenamefont {Day}}]{Gao07}%
  \BibitemOpen
  \bibfield  {author} {\bibinfo {author} {\bibfnamefont {J.}~\bibnamefont
  {Gao}}, \bibinfo {author} {\bibfnamefont {J.}~\bibnamefont {Zmuidzinas}},
  \bibinfo {author} {\bibfnamefont {B.}~\bibnamefont {Mazin}}, \bibinfo
  {author} {\bibfnamefont {H.}~\bibnamefont {LeDuc}}, \ and\ \bibinfo {author}
  {\bibfnamefont {P.~K.}\ \bibnamefont {Day}},\ }\href@noop {} {\bibfield
  {journal} {\bibinfo  {journal} {Appl. Phys. Lett.}\ }\textbf {\bibinfo
  {volume} {90}},\ \bibinfo {pages} {102507} (\bibinfo {year}
  {2007})}\BibitemShut {NoStop}%
\bibitem [{\citenamefont {Burnett}\ \emph
  {et~al.}(2014{\natexlab{a}})\citenamefont {Burnett}, \citenamefont {Faoro},
  \citenamefont {Wisby}, \citenamefont {Gurtovoi}, \citenamefont {Chernykh},
  \citenamefont {Mikhailov}, \citenamefont {Tulin}, \citenamefont
  {Shaikhaidarov}, \citenamefont {Antonov}, \citenamefont {Meeson},
  \citenamefont {Tzalenchuk},\ and\ \citenamefont
  {Lindstr\"om}}]{FaoroResonators}%
  \BibitemOpen
  \bibfield  {author} {\bibinfo {author} {\bibfnamefont {J.}~\bibnamefont
  {Burnett}}, \bibinfo {author} {\bibfnamefont {L.}~\bibnamefont {Faoro}},
  \bibinfo {author} {\bibfnamefont {I.}~\bibnamefont {Wisby}}, \bibinfo
  {author} {\bibfnamefont {V.}~\bibnamefont {Gurtovoi}}, \bibinfo {author}
  {\bibfnamefont {A.~V.}\ \bibnamefont {Chernykh}}, \bibinfo {author}
  {\bibfnamefont {G.}~\bibnamefont {Mikhailov}}, \bibinfo {author}
  {\bibfnamefont {V.}~\bibnamefont {Tulin}}, \bibinfo {author} {\bibfnamefont
  {R.}~\bibnamefont {Shaikhaidarov}}, \bibinfo {author} {\bibfnamefont
  {V.}~\bibnamefont {Antonov}}, \bibinfo {author} {\bibfnamefont
  {P.}~\bibnamefont {Meeson}}, \bibinfo {author} {\bibfnamefont {A.~Y.}\
  \bibnamefont {Tzalenchuk}}, \ and\ \bibinfo {author} {\bibfnamefont
  {T.}~\bibnamefont {Lindstr\"om}},\ }\href@noop {} {\bibfield  {journal}
  {\bibinfo  {journal} {Nat. Commun.}\ }\textbf {\bibinfo {volume} {5}},\
  \bibinfo {pages} {4119} (\bibinfo {year} {2014}{\natexlab{a}})}\BibitemShut
  {NoStop}%
\bibitem [{\citenamefont {Sage}\ \emph {et~al.}(2011)\citenamefont {Sage},
  \citenamefont {Bolkhovsky}, \citenamefont {Oliver}, \citenamefont {Turek},\
  and\ \citenamefont {Welander}}]{TLSloss}%
  \BibitemOpen
  \bibfield  {author} {\bibinfo {author} {\bibfnamefont {J.~M.}\ \bibnamefont
  {Sage}}, \bibinfo {author} {\bibfnamefont {V.}~\bibnamefont {Bolkhovsky}},
  \bibinfo {author} {\bibfnamefont {W.~D.}\ \bibnamefont {Oliver}}, \bibinfo
  {author} {\bibfnamefont {B.}~\bibnamefont {Turek}}, \ and\ \bibinfo {author}
  {\bibfnamefont {P.~B.}\ \bibnamefont {Welander}},\ }\href {\doibase
  10.1063/1.3552890} {\bibfield  {journal} {\bibinfo  {journal} {J. Appl.
  Phys.}\ }\textbf {\bibinfo {volume} {109}},\ \bibinfo {pages} {063915}
  (\bibinfo {year} {2011})}\BibitemShut {NoStop}%
\bibitem [{\citenamefont {Wang}\ \emph {et~al.}(2009)\citenamefont {Wang},
  \citenamefont {Hofheinz}, \citenamefont {Wenner}, \citenamefont {Ansmann},
  \citenamefont {Bialczak}, \citenamefont {Lenander}, \citenamefont {Lucero},
  \citenamefont {Neeley}, \citenamefont {O'Connell}, \citenamefont {Sank},
  \citenamefont {Weides}, \citenamefont {Cleland},\ and\ \citenamefont
  {Martinis}}]{TLSloss2}%
  \BibitemOpen
  \bibfield  {author} {\bibinfo {author} {\bibfnamefont {H.}~\bibnamefont
  {Wang}}, \bibinfo {author} {\bibfnamefont {M.}~\bibnamefont {Hofheinz}},
  \bibinfo {author} {\bibfnamefont {J.}~\bibnamefont {Wenner}}, \bibinfo
  {author} {\bibfnamefont {M.}~\bibnamefont {Ansmann}}, \bibinfo {author}
  {\bibfnamefont {R.~C.}\ \bibnamefont {Bialczak}}, \bibinfo {author}
  {\bibfnamefont {M.}~\bibnamefont {Lenander}}, \bibinfo {author}
  {\bibfnamefont {E.}~\bibnamefont {Lucero}}, \bibinfo {author} {\bibfnamefont
  {M.}~\bibnamefont {Neeley}}, \bibinfo {author} {\bibfnamefont {A.~D.}\
  \bibnamefont {O'Connell}}, \bibinfo {author} {\bibfnamefont {D.}~\bibnamefont
  {Sank}}, \bibinfo {author} {\bibfnamefont {M.}~\bibnamefont {Weides}},
  \bibinfo {author} {\bibfnamefont {A.~N.}\ \bibnamefont {Cleland}}, \ and\
  \bibinfo {author} {\bibfnamefont {J.~M.}\ \bibnamefont {Martinis}},\ }\href
  {\doibase 10.1063/1.3273372} {\bibfield  {journal} {\bibinfo  {journal}
  {Appl. Phys. Lett.}\ }\textbf {\bibinfo {volume} {95}},\ \bibinfo {pages}
  {233508} (\bibinfo {year} {2009})}\BibitemShut {NoStop}%
\bibitem [{\citenamefont {Goetz}\ \emph {et~al.}(2016)\citenamefont {Goetz},
  \citenamefont {Deppe}, \citenamefont {Haeberlein}, \citenamefont {Wulschner},
  \citenamefont {Zollitsch}, \citenamefont {Meier}, \citenamefont {Fischer},
  \citenamefont {Eder}, \citenamefont {Xie}, \citenamefont {Fedorov},
  \citenamefont {Menzel}, \citenamefont {Marx},\ and\ \citenamefont
  {Gross}}]{TLSloss3}%
  \BibitemOpen
  \bibfield  {author} {\bibinfo {author} {\bibfnamefont {J.}~\bibnamefont
  {Goetz}}, \bibinfo {author} {\bibfnamefont {F.}~\bibnamefont {Deppe}},
  \bibinfo {author} {\bibfnamefont {M.}~\bibnamefont {Haeberlein}}, \bibinfo
  {author} {\bibfnamefont {F.}~\bibnamefont {Wulschner}}, \bibinfo {author}
  {\bibfnamefont {C.~W.}\ \bibnamefont {Zollitsch}}, \bibinfo {author}
  {\bibfnamefont {S.}~\bibnamefont {Meier}}, \bibinfo {author} {\bibfnamefont
  {M.}~\bibnamefont {Fischer}}, \bibinfo {author} {\bibfnamefont
  {P.}~\bibnamefont {Eder}}, \bibinfo {author} {\bibfnamefont {E.}~\bibnamefont
  {Xie}}, \bibinfo {author} {\bibfnamefont {K.~G.}\ \bibnamefont {Fedorov}},
  \bibinfo {author} {\bibfnamefont {E.~P.}\ \bibnamefont {Menzel}}, \bibinfo
  {author} {\bibfnamefont {A.}~\bibnamefont {Marx}}, \ and\ \bibinfo {author}
  {\bibfnamefont {R.}~\bibnamefont {Gross}},\ }\href {\doibase
  http://dx.doi.org/10.1063/1.4939299} {\bibfield  {journal} {\bibinfo
  {journal} {J. Appl. Phys.}\ }\textbf {\bibinfo {volume} {119}},\ \bibinfo
  {pages} {015304} (\bibinfo {year} {2016})}\BibitemShut {NoStop}%
\bibitem [{\citenamefont {Schickfus}\ and\ \citenamefont
  {Hunklinger}(1977)}]{Schickfus77}%
  \BibitemOpen
  \bibfield  {author} {\bibinfo {author} {\bibfnamefont {M.~V.}\ \bibnamefont
  {Schickfus}}\ and\ \bibinfo {author} {\bibfnamefont {S.}~\bibnamefont
  {Hunklinger}},\ }\href {\doibase
  http://dx.doi.org/10.1016/0375-9601(77)90558-8} {\bibfield  {journal}
  {\bibinfo  {journal} {Phys. Lett. A}\ }\textbf {\bibinfo {volume} {64}},\
  \bibinfo {pages} {144 } (\bibinfo {year} {1977})}\BibitemShut {NoStop}%
\bibitem [{\citenamefont {Burnett}\ \emph
  {et~al.}(2014{\natexlab{b}})\citenamefont {Burnett}, \citenamefont {Faoro},
  \citenamefont {Wisby}, \citenamefont {Gurtovoi}, \citenamefont {Chernykh},
  \citenamefont {Mikhailov}, \citenamefont {Tulin}, \citenamefont
  {Shaikhaidarov}, \citenamefont {Antonov}, \citenamefont {Meeson},
  \citenamefont {Tzalenchuk},\ and\ \citenamefont
  {Lindstr\"om}}]{burnett_evidence_2014}%
  \BibitemOpen
  \bibfield  {author} {\bibinfo {author} {\bibfnamefont {J.}~\bibnamefont
  {Burnett}}, \bibinfo {author} {\bibfnamefont {L.}~\bibnamefont {Faoro}},
  \bibinfo {author} {\bibfnamefont {I.}~\bibnamefont {Wisby}}, \bibinfo
  {author} {\bibfnamefont {V.~L.}\ \bibnamefont {Gurtovoi}}, \bibinfo {author}
  {\bibfnamefont {A.~V.}\ \bibnamefont {Chernykh}}, \bibinfo {author}
  {\bibfnamefont {G.~M.}\ \bibnamefont {Mikhailov}}, \bibinfo {author}
  {\bibfnamefont {V.~A.}\ \bibnamefont {Tulin}}, \bibinfo {author}
  {\bibfnamefont {R.}~\bibnamefont {Shaikhaidarov}}, \bibinfo {author}
  {\bibfnamefont {V.}~\bibnamefont {Antonov}}, \bibinfo {author} {\bibfnamefont
  {P.~J.}\ \bibnamefont {Meeson}}, \bibinfo {author} {\bibfnamefont {A.~Y.}\
  \bibnamefont {Tzalenchuk}}, \ and\ \bibinfo {author} {\bibfnamefont
  {T.}~\bibnamefont {Lindstr\"om}},\ }\href
  {http://dx.doi.org/10.1038/ncomms5119} {\bibfield  {journal} {\bibinfo
  {journal} {Nat. Comm.}\ }\textbf {\bibinfo {volume} {5}},\ \bibinfo {pages}
  {4119} (\bibinfo {year} {2014}{\natexlab{b}})}\BibitemShut {NoStop}%
\bibitem [{\citenamefont {Ramanayaka}\ \emph {et~al.}(2015)\citenamefont
  {Ramanayaka}, \citenamefont {Sarabi},\ and\ \citenamefont
  {Osborn}}]{1/fnoise}%
  \BibitemOpen
  \bibfield  {author} {\bibinfo {author} {\bibfnamefont {A.~N.}\ \bibnamefont
  {Ramanayaka}}, \bibinfo {author} {\bibfnamefont {B.}~\bibnamefont {Sarabi}},
  \ and\ \bibinfo {author} {\bibfnamefont {K.~D.}\ \bibnamefont {Osborn}},\
  }\href@noop {} {\bibfield  {journal} {\bibinfo  {journal} {arXiv}\ }
  (\bibinfo {year} {2015})},\ \Eprint {http://arxiv.org/abs/1507.06043}
  {arXiv:1507.06043 [cond-mat.supr-con]} \BibitemShut {NoStop}%
\bibitem [{\citenamefont {Burin}\ \emph {et~al.}(2015)\citenamefont {Burin},
  \citenamefont {Matityahu},\ and\ \citenamefont {Schechter}}]{1/fnoise_burin}%
  \BibitemOpen
  \bibfield  {author} {\bibinfo {author} {\bibfnamefont {A.~L.}\ \bibnamefont
  {Burin}}, \bibinfo {author} {\bibfnamefont {S.}~\bibnamefont {Matityahu}}, \
  and\ \bibinfo {author} {\bibfnamefont {M.}~\bibnamefont {Schechter}},\ }\href
  {\doibase 10.1103/PhysRevB.92.174201} {\bibfield  {journal} {\bibinfo
  {journal} {Phys. Rev. B}\ }\textbf {\bibinfo {volume} {922}},\ \bibinfo
  {pages} {174201} (\bibinfo {year} {2015})}\BibitemShut {NoStop}%
\bibitem [{\citenamefont {Faoro}\ and\ \citenamefont
  {Ioffe}(2015)}]{1/fnoise_faoro}%
  \BibitemOpen
  \bibfield  {author} {\bibinfo {author} {\bibfnamefont {L.}~\bibnamefont
  {Faoro}}\ and\ \bibinfo {author} {\bibfnamefont {L.~B.}\ \bibnamefont
  {Ioffe}},\ }\href {\doibase 10.1103/PhysRevB.91.014201} {\bibfield  {journal}
  {\bibinfo  {journal} {Phys. Rev. B}\ }\textbf {\bibinfo {volume} {91}},\
  \bibinfo {pages} {014201} (\bibinfo {year} {2015})}\BibitemShut {NoStop}%
\end{thebibliography}%

\end{document}